\documentclass[aps,prd,amsmath,amssymb]{revtex4}
\usepackage{graphicx}
\usepackage{bm}

\def\journal#1#2#3#4{{#1} {\bf #2}, #3 (#4)}
\newcommand{\be}{\begin{equation}}
\newcommand{\ee}{\end{equation}}
\newcommand{\bea}{\begin{eqnarray}}
\newcommand{\eea}{\end{eqnarray}}
\newcommand{\hf}{\frac12}
\newcommand{\nn}{\nonumber\\}
\def\eq#1{(\ref{#1})}

\def\ord#1{{\cal O}\left(#1\right)}
\def\mr#1{{\mathrm{#1}}}
\def\v#1{{\bm{#1}}}
\def\fd#1#2{\frac{\delta#1}{\delta#2}}
\def\fdd#1#2#3{\frac{\delta^2#1}{\delta#2\delta#3}}
\def\dt{{\Delta t}}
\def\hj{{\hat j}}
\def\hphi{\hat\phi}
\def\hx{{\hat x}}
\def\hA{{\hat A}}
\def\hD{{\hat D}}

\def\Im{\mr{Im}}
\def\Re{\mr{Re}}

\begin{document}
\title{Effective dynamics of a classical point charge}
\author{Janos Polonyi}
\email{polonyi@iphc.cnrs.fr}
\affiliation{Strasbourg University, CNRS-IPHC, 23 rue du Loess, BP28 67037 Strasbourg Cedex 2 France}
\date{\today}

\begin{abstract}
The effective Lagrangian of a point charge is derived by eliminating the electromagnetic field within the framework of the classical closed time path formalism. The short distance singularity of the electromagnetic field is regulated by an UV cutoff. The Abraham-Lorentz force is recovered and its similarity to quantum anomalies is underlined. The full cutoff-dependent linearized equation of motion is obtained, no runaway trajectories are found but the effective dynamics shows acausality if the cutoff is beyond the classical charge radius. The strength of the radiation reaction force displays a pole in its cutoff-dependence in a manner reminiscent of the Landau-pole of perturbative QED. Similarity between the dynamical breakdown of the time reversal invariance and dynamical symmetry breaking is pointed out.
\end{abstract}
\maketitle

\section{Introduction}
Classical electrodynamics of point charges has an intrinsic length scale, the classical charge radius $r_0=e^2/mc^2$, separating the well known classical domain from an unusual classical world, hidden behind quantum effects. Nevertheless it is an intriguing question how this classical world of point charges would look like in the absence of quantum mechanics. The attempts to uncover the dynamics around or beyond to the crossover scale, $r_0$, are rendered difficult by the current state of understanding of the radiation reaction force \cite{brl}, an important ingredient of the electromagnetic interaction at these scales. The problem comes from the singularity of the electromagnetic field (EMF) in the point-like  limit of the charges and appears as an instability, generated by the Abraham-Lorentz force, the runaway solution \cite{dirac}.

A simple, systematic derivation of the radiation reaction is presented here by working out the effective dynamics of a single point charge. The singularities of the point charge limit are regulated by smearing the interactions within an invariant length $\ell$, an UV cutoff,  and the dissipative nature of the radiation reaction force is handled by an extension of the variational method of classical mechanics, by recasting Schwinger's Closed Time Path (CTP) formalism \cite{schw}. Though this scheme has already been applied in a number of problems its advantages and power have not yet been exhausted. It has been used to find relaxation in many-body quantum systems \cite{kadanoffb}, to generate perturbation expansion for retarded Green functions in quantum field theory \cite{keldysh}, to find manifestly time reversal invariant description of quantum mechanics \cite{aharonov}, to describe finite temperature effects in quantum field theory \cite{umezawa}, to find the mixed state contributions to the density matrix by the path integral \cite{feynmanv}, to describe non-equilibrium processes \cite{calzetta}, to derive equations of motion for the expectation value of local operators \cite{jo,ed}, to describe scattering processes with non-equilibrium final states \cite{scatt} and finally, to derive radiation reaction force in QED \cite{johnson,galleyh,galleyl} and effective quantum gravity \cite{galleyt}. We use it in this paper without referring to quantum theory, by generalizing the variational principle of classical mechanics \cite{bateman,clctp,galley}. This generalization provides a compromise which on the one hand, preserves the functional formalism of the variational principle, a natural way to introduce Green functions and perturbation expansion, and on the other hand, it allows the imposition of initial conditions and can handle dissipative forces.

The well known expression of the Abraham-Lorentz force is recovered in the calculation, together with the usual, UV divergent mass renormalization in the limit $\ell\to0$. The linearized nonlocal equation of motion is derived for a point charge by retaining the full cutoff dependence. It is pointed out that we do not possess all data necessary to solve an initial condition problem and the solution is constructed in terms of the retarded Green function. The mode which drives the usual runaway solution is absent and the motion is causal for $\ell\gg r_0$. For sufficiently small values of $\ell$ unstable modes appear but they do not lead to runaway solutions because the retarded Green function relies on the unstable modes in the acausal regime where these modes are bounded. The elimination of the EMF generates UV singularities, in a manner similar to quantum field theories. This requires the introduction of an UV cutoff, a shift of the radiation field modes off the light-cone. It is pointed out that the integral over the world-line of the charges, contributing to the effective action is not uniformly convergent in the limit when the cutoff is removed, in a manner reminiscent of anomalies in quantum field theory. This similarity is made more explicit by showing that the EMF remains slightly off light-cone even after the removal of the cutoff.

The organization of this paper is the following. Section \ref{ctps} introduces the CTP formalism in classical field theory which gives a systematic definition of the retarded Green function in classical effective theories. The way the time arrow is generated by the environment, irreversibility, acausality arise and the runaway solution is avoided are the subject of Section \ref{timearrs}. Section \ref{pcefths} contains the derivation of the effective equation of motion. Finally, the summary of the results is presented in Section \ref{conls}. Some details about the CTP Green function are collected in Appendix \ref{ctpgrfcap}.

\section{CTP}\label{ctps}
An extension of the variational method of classical mechanics is needed to cover dissipative forces in a functional framework where the solution of initial condition problems can be found by means of a systematically derived retarded Green function. We consider a classical system described by the coordinate $x$, governed by the Lagrangian $L(x,\dot x)$. A solution of the equation of motion for $t_i\le t\le t_f$ is usually identified by imposing auxiliary conditions, such as $x(t_i)=x_i$ and $x(t_f)=x_f$. Can we replace these boundary conditions with the initial conditions $x(t_i)=x_i$, $\dot x(t_i)=v_i$? The problem is that the equation of motion must then be imposed at the final time and it cancels the generalized momentum. This condition can be avoided by constructing a particular trajectory, $\tilde x(\tilde t)$, for twice as long time interval, $0<\tilde t<2(t_f-t_i)$ which satisfies the desired initial conditions. The particularity of the trajectory is that a time reversal transformation is performed at $\tilde t=t_f-t_i$, first the change $\dot{\tilde x}(t_f-t_i)\to-\dot{\tilde x}(t_f-t_i)$ is performed and then the same equation of motion is solved backward in time. The motion stops at $\tilde t=2(t_f-t_i)$ when $\tilde x$ arrives at the time reversed initial conditions. A more convenient book-keeping which is used below consists of a formal reduplication of the degrees of freedom, $\tilde x\to\hat x=(x_+,x_-)$ where the members of the CTP doublet are defined for $t_i\le t\le  t_f$ as $x_+(t)=\tilde x(t-t_i)$ and $x_-(t)=\tilde x(2t_f-t_i-t)$.

It will be important to distinguish true time reversal transformation from a formal reparametrization of the motion because the auxiliary conditions are handled differently. Both involve the reversal of the direction of time in the equation of motion and the exchange of the initial and final time, $t_i\leftrightarrow t_f$, in the auxiliary conditions. The values of $x_i$ and $v_i$ are transformed as $x_i\to x_f$ and $v_i\to-v_f$ in time reversal and the solution changes if the equation of motion is not time reversal invariant. In the time reversed reparametrization the values of $x_i$ and $v_i$ are adjusted to recover the same trajectory in reversed time, irrespectively of the time reversal properties of the equation of motion.

\subsection{Finite time motion}\label{ctpaft}
The variational principle is based on the action
\be\label{ctpact}
S_{CTP}[\hat x]=\int_{t_i}^{t_f}dt[L_{CTP}(x_+(t),\dot x_+(t))-L_{CTP}^*(x_-(t),\dot x_-(t))],
\ee
where $\Re L_{CTP}=L$ is the original Lagrangian. The variation is within the set of trajectories defined by the auxiliary conditions, namely both members of the CTP doublet satisfy the initial conditions
\be\label{ctpinc}
x_\pm(t_i)=x_i,~~~\dot x_\pm(t_i)=v_i,
\ee
and the constraint
\be\label{ctpfin}
x_+(t_f)=x_-(t_f).
\ee
This latter assures that the boundary term, arising in the calculation of the equation of motion from the variation at the final time cancels. The opposite signs in front of the two Lagrangians are to render the variational equation trivial, $0=0$, at the final time as far as the real part of the action is concerned. If the original Lagrangian were used for both trajectories then the CTP action would degenerate for the CTP doublets $x_+(t)=x_-(t)$. This degeneracy can be lifted by introducing an infinitesimal difference between the forward and backward running dynamics in time. The simplest difference, introduced between the two time axes which  switches off adiabatically the auxiliary conditions in the limit $t_f-t_i\to\infty$ as expected corresponds to
\be\label{ctplag}
L_{CTP}(x,\dot x)=L(x,\dot x)+\frac{i\epsilon}2x^2
\ee
and the limit $\epsilon\to0$ is supposed to be performed after deriving and solving the variational equations of motion as in Feynman $i\epsilon$ prescription in the quantum case. Note that the formal reparametrization of the motion, called CTP transformation below, generates $S_{CTP}\to-S^*_{CTP}$ and preserves the equation of motion.

\subsection{Green function}\label{ctpgfvs}
To find the Green function we consider a harmonic system, defined by the action
\be\label{ctpquact}
S_{CTP}[\hat x]=\int dt\left[\hf\hat x(t)\hat K\hat x(t)+\hj(t)\hat x(t)\right],
\ee
and the CTP Green function, $\hD=\hat K^{-1}$ yields the trajectory
\be\label{ctpgfv}
\hat x(t)=-\int dt'\hD(t,t')\hj(t'),
\ee
written as $\hat x=-\hD\hj$ in condensed notation. The $\ord{\epsilon}$ term in the CTP action makes the null-space of $\hat K$ trivial and its inverse, $\hD$, well defined.

The CTP transformation, $S_{CTP}\to-S^*_{CTP}$, under time reversed reparametrization requires the block structure
\be\label{quadrfa}
\hat K=\hat\sigma\begin{pmatrix}K_n+iK_{i,1}&-K_f+iK_{i,2}\cr K_f+iK_{i,2}&-K_n+iK_{i,1}\end{pmatrix}\hat\sigma,
\ee
where $K_{i,1}$, $K_{i,2}$, $K_n$ and $K_f$ are real functions and
\be
\hat\sigma=\begin{pmatrix}1&0\cr0&-1\end{pmatrix}.
\ee
denotes the CTP ``metric tensor''.It is easy to see that the inverse has a similar block structure,
\be\label{trevgrfv}
\hD=\begin{pmatrix}D_n+iD_{i,1}&-D_f+iD_{i,2}\cr D_f+iD_{i,2}&-D_n+iD_{i,1}\end{pmatrix}.
\ee
including four real functions $D_{i,1}$, $D_{i,2}$, $D_n$ and $D_f$. The quadratic form \eq{quadrfa} and its inverse \eq{trevgrfv} are symmetric, $\hD^{tr}=\hD$, therefore the relations $D_{i,1}^{tr}=D_{i,1}$, $\tilde D_{i,2}^{tr}=D_{i,2}$, $\tilde D_n^{tr}=D_n$, $D_f^{tr}=-D_f$ follow.

Though we introduced two independent sources, $\hj=(j_+,j_-)$ to diagnose the forward and backward moving parts of the dynamics independently, the physical external source corresponds to the particular choice $\hj=(j,-j)$. The solution of the equation of motion can be written as
\be\label{phtraj}
x_\sigma(t)=-\sum_{\sigma'=\pm1}\int_{t_i}^{t_f}dt'D_{\sigma\sigma'}(t,t')\sigma'j(t')
\ee
in this case and to find a (no-hat) $\sigma$-independent solution we need the identity
\be\label{ctpstr}
D_{++}+D_{--}=D_{-+}+D_{+-}
\ee
which imposes $D_{i,1}=D_{i,2}$,
\be\label{ctpstrgrfv}
\hD=\begin{pmatrix}D_n+iD_i&-D_f+iD_i\cr D_f+iD_i&-D_n+iD_i\end{pmatrix}.
\ee

The trajectory \eq{phtraj} suggests the introduction of the retarded Green function $D_r=D_n+D_f$, and the advanced Green function is found by time reversal transformation, $D_a=D_n-D_f$. The EMF strength, $F_{\mu\nu}$, can be split into near and far field components according to their decrease with the distance from the source. This splitting can be carried over the vector potential $A_\mu$, as well. Though the distance dependence is not a reliable condition for a gauge dependent field but the splitting can be made well defined by noting that the Green functions which give those components of the Li\`enard-Wiecher potential which produce the near and far field strengths are the time reversal symmetric ($D_n$) and antisymmetric ($D_f$) parts of the retarded Green function, respectively \cite{dirac}. A further justification of the choice of $D_f=(D_r-D_a)/2$, based on the relative minus sign in the CTP Lagrangian \eq{ctpact} will be given below, in Section \ref{ctpeaed}.

Causality, the effect of an external perturbation appearing after the time of the action of the perturbation further reduces the number of independent functions and requires $D_f(t,t')=\mr{sign}(t-t')D_n(t,t')$. It is advantageous to define the retarded and advanced combinations of the quadratic form of the action, $K_{\stackrel{r}{a}}=K_n\pm K_f$ and elementary algebra gives the relations  \cite{maxwell}
\be\label{invprra}
K_{\stackrel{r}{a}}=D_{\stackrel{r}{a}}^{-1},
\ee
and $K_{i,1}=K_{i,2}=K_i$ with
\be\label{invpri}
K_i=-D_r^{-1}D_iD_a^{-1}.
\ee
The choice of the imaginary part of the action remains free in classical physics for this part of the Green function controls quantum fluctuations only. The explicit calculation of the Green function is outlined in Appendix \ref{ctpgrfcap}. Time reversal flips the direction of the time in which the external source is visited during the motion and it is represented by the transformation $\hD(x,x')\to\hD(Tx,Tx')$, where $T(t,\v{x})=(-t,\v{x})$ and induces $D_f\to-D_f$,  $D_r\leftrightarrow D_a$ and $K_f\to-K_f$.

\subsection{Infinite time motion}\label{infttmot}
The action \eq{ctpact} and the auxiliary conditions describe the dynamics in a finite time interval and its Green function is not invariant under translations in time. The extension of the action principle is not obvious for $t_f-t_i\to\infty$ because the auxiliary condition \eq{ctpfin} must be preserved or else the two members of the CTP doublets decouple. But the action can be constructed in this limit by first calculating the Green function for $t_f-t_i\to\infty$ and defining the quadratic form of the action as the inverse of the Green function. Let us consider a free scalar field, described by the action
\be\label{ctpfa}
S_{CTP}[\hphi]=\int d^4x\left[\hf\hphi(x)\hat K\hat\phi(x)+\hj(x)\hphi(x)\right].
\ee
The Green function is calculated in Appendix \ref{ctpgrfcap} and its inverse, given by Eq. \eq{ifreectpgfv} leads to
\be\label{inttctpa}
S_{CTP}[\hphi]=\hf\int\frac{d\omega d^3p}{(2\pi)^4}(\phi_+(-\omega,-\v{p}),\phi_-(-\omega,-\v{p}))\begin{pmatrix}\omega^2-\v{p}^2-m^2+i\epsilon&-2i\epsilon\Theta(-\omega)\cr-2i\epsilon\Theta(\omega)&-\omega^2+\v{p}^2+m^2+i\epsilon\end{pmatrix}\begin{pmatrix}\phi_+(\omega,\v{p})\cr\phi_-(\omega,\v{p})\end{pmatrix},
\ee
where $p^2=p^{02}-\v{p}^2$ and $K_n=p^2-m^2$, $K_f=i\mr{sign}(p^0)\epsilon$ and $K_i=\epsilon$.

Few remarks are in order at this point. (i) Translation invariance in time is recovered in \eq{inttctpa} by the adiabatic switching off the auxiliary conditions by the imaginary part of the action \eq{ctplag}. (ii) Note the transmutation of the coupling between the two time axes: the identification of the final coordinates of the two members of the CTP doublet for the finite $t_f-t_i$ is traded into a coupling of infinitesimal, $\ord{\epsilon}$ strength for $-\infty<t<\infty$ in the action \eq{inttctpa}. (iii) The trajectory, given by the retarded Green function in the limit $t_f-t_i\to\infty$ corresponds to the trivial, $x_i=v_i=0$ initial conditions, imposed in the distant past.

\subsection{Effective theory}
The effective system dynamics is generated by the elimination of the environment variables by means of their equations of motion and auxiliary conditions. This procedure is now followed in a functional formalism which allows us to identify the retarded Green function in a systematic manner, without adjusting the pole structure by hand.

Let us introduce the system and environment coordinates, $x$ and $y$, respectively and write the original multi-component coordinate as $X=X(x,y)$. The effective system dynamics is found by following the usual procedure of quantum field theory: one introduces a book-keeping source, $j$, and constructs the Legendre transform
\be\label{genfcgfv}
W[j]=S[x,y]+\int dtj(t)x(t),
\ee
where $S[x,y]=S[X(x,y)]$ and $y$ satisfies the equation of motion $\delta S[x,y]/\delta y=0$, equipped by some auxiliary conditions which make the solution unique. As a result, the equation $\delta W[j]/\delta j=x$ follows and the effective action is defined by the inverse Legendre transform,
\be\label{effact}
S_{eff}[x]=W[j]-\int dtj(t)x(t).
\ee
The relation
\be\label{neqm}
\fd{S_{eff}[x]}{x}=-j,
\ee
suggests the interpretation of $S_{eff}[x]$ as effective system action. The construction makes it evident that this elimination is achieved by solving the equation of motion for the environment coordinates in \eq{genfcgfv} and recalculating the action in \eq{effact} in terms of the system coordinates. The advantage of the functional definition of the  effective action is the simple definition of the effective Green function,
\be\label{gffd}
D=\left(\fdd{S_{eff}[0]}{x}{x}\right)^{-1}.
\ee
The same procedure can be repeated within the CTP formalism where the CTP symmetry allows us to write the effective action as
\be\label{ctpeffact}
S_{eff}[\hat x]=S_0[x_+]-S_0^*[x_-]+S_I[\hat x].
\ee
The definition of the influence functional $S_I$ \cite{feynmanv}, the coupling the two time axes is rendered unique by imposing $\delta^2S_I[x_+,x_-]/\delta x_+\delta x_-\ne0$.

It is illuminating to separate the system-environment interactions into conserving and non-conserving classes. We talk about the former if the system energy-momentum conservation can be recovered by a suitable redefinition, ``renormalization`` of the system energy-momentum. In a more realistic many-body system this happens when the system-environment interaction is localized in space-time, for instance in the case of a polaron in solids. The contributions of conserving interactions are generated by the near field Green function and appear in the single time-axes piece of the effective action, $S_0$, reflecting the conservative nature of such a ''dressing``. The conserving interactions are described by the exchange of a virtual particle in quantum field theory models. Non-conserving interactions induce asymptotically long time energy increase in the environment, realized by delocalized excitations. They are represented by the far field Green function and the exchange of real, mass-sell particles in quantum field theory. Their contributions appear in the influence functional, $S_I$. A distinguishing feature of the CTP formalism is the explicit separation of these two kinds of interactions and the mapping of non-conserving interactions into interactions between two members of the CTP doublets, representing the same physical system.

\section{Time arrow}\label{timearrs}
We address here the breakdown of time reversal symmetry and the causal structure furthermore the fate of unbounded, runaway trajectories in effective theories. The effective dynamics where the time reversal invariance is broken dynamically by the environment is irreversible because dynamical symmetry breaking implies infinitely many weakly coupled environment degrees of freedom and allows the application of the methods of statistical physics. Causality is a logical relation, namely A is the cause of B if the appearance of A always leads to B. The runaway modes will be sought in harmonic systems where they grow exponentially in time. All these are related to the time arrow.

A possible experimental determination of the time arrow is the application of an external perturbation which is localized in time: the time arrow is forward or backward if the response is found after or before the perturbation, respectively. The time arrow, established in this manner is relative, it is defined with respect to the flow of the proper time of the observer. The time arrow is set in our calculations of the trajectory of a local equation of motion by imposing initial or final auxiliary conditions. One may go in thought beyond experimental limitations and diagnose the calculated trajectory either by imposing opposite time arrows on different degrees of freedom or seek the dynamical generation of time arrows, independent of auxiliary conditions.

\subsection{Time reversal of the effective dynamics}
The distinguishing feature of time inversion, compared to other space-time symmetry transformations is the important role of auxiliary conditions in breaking time reversal invariance. Though this is an obvious issue for closed, autonomous dynamics, the auxiliary conditions of the undetected environment make the time reversal invariance of the effective system dynamics a non-trivial issue. Let us follow this problem in a simple scheme where the system and environment coordinates are denoted by $x$ and $y$, respectively, and they obey time reversal and time translation invariant equations of motion,
\be\label{seqm}
\ddot x=F(x,y),~~~\ddot y=G(x,y),
\ee
with separable initial conditions, $f(x(t_i),\dot x(t_i))=g(y(t_i),\dot y(t_i))=0$. The effective system dynamics is constructed by first solving the environment equation of motion for an arbitrary system trajectory, $y=y[x,g]$, and after that inserting this solution into the system equation of motion,
\be\label{efemgm}
\ddot x=F(x,y[x,g]).
\ee

Time reversal and time translations are realized in a nontrivial manner by the effective equation of motion due to the presence of the environment auxiliary condition $g(y)$. The time translation invariance can be recovered if the environment is large enough, namely if generates a relaxation process toward an equilibrium state. We perform in this case first the thermodynamical limit, $N\to\infty$ where $N$ denotes the number of weakly coupled environmental degrees of freedom and after that the limit $t_f-t_i\to\infty$. The second, long time limit decouples the environment auxiliary conditions and restores time translation invariance, but the fate of time reversal invariance remains open. The effective dynamics is called irreversible if its effective equation of motion recovers time translation invariance but the environment auxiliary conditions violate the formal time reversal invariance of the effective equation of motion. The solution of an irreversible equation of motion displays relaxation and dissipative phenomena in one direction of the time and becomes unstable and may diverge if the time runs in the opposite direction.

We have no empirical basis for absolute time arrow or causality in the absence of a ``control Universe'' to compare phenomena in the presence or absence of an external perturbation. But relaxation to an equilibrium and irreversibility represents a practical approximation of the complicated system dynamics where the time arrow becomes absolute. It is based on the double limit $N\to\infty$ and $t_f-t_i\to\infty$ which allows a stable time evolution in one direction of the time only and remains applicable up to time scales far beyond reasonable observation length.

The emergence of irreversibility is demonstrated below in the framework of a simple toy model.

\subsection{Toy model}\label{toymod}
A simple but instructive toy model can be made for the system and environment coordinates, $x$ and $y_n$, $n=1,\ldots,N$, respectively, by introducing the Lagrangian \cite{hamod}
\be\label{toym}
L=\frac{m}{2}\dot x^2-\frac{m\omega^2_0}2x^2+jx+\sum_n\left(\frac{m}2\dot y^2_n-\frac{m\omega_n^2}2y^2_n-g_ny_nx\right),
\ee
where $\omega_n\ge0$ and the bound
\be\label{stab}
\sum_n\frac{g_n^2}{\omega^2_n}<m^2\omega_0^2
\ee
is required to have bounded potential energy and real frequency spectrum. We use the time arrow $\tau=1$ or $\tau=-1$ to indicate that the trivial auxiliary conditions, i.e. vanishing coordinate and velocity are specified at $t=t_i$ or $t=t_f$, respectively. All environment degrees of freedom have the same time arrow. The effective system equation of motion,
\be\label{efseqm}
-j(\omega)=[m(\omega^2-\omega^2_0)-\Sigma_{\tau_e}(\omega)]x(\omega),
\ee
contains the self energy,
\be
\Sigma_{\tau_e}(\omega)=\sum_n\frac{g^2_n}{m}\frac1{(\omega+i\tau_e\epsilon)^2-\omega_n^2},
\ee
and the retarded system Green function is given by
\be\label{grfsw}
D_{\tau_s,\tau_e}(t)=\int\frac{d\omega}{2\pi}\frac{e^{-i\omega t}}{m[(\omega+i\tau_s\epsilon)^2-\omega_0^2]-\Sigma_{\tau_e}(\omega)}.
\ee
It is easy to see that the poles of the Green function of discrete spectrum have infinitesimal negative imaginary part for consistent time arrows, $\tau_s=\tau_e$, making the effective dynamics reversible and causal. Conflicting time arrows, $\tau_s=-\tau_e$,  may induce acausality.

The environment is not monitored in an effective theory where the time reversal consists of the transformation $(\tau_s,\tau_e)\to(-\tau_s,\tau_e)$. The effective equation of motion, $D^{-1}x=0$, contains even powers of the time derivatives and displays formal time reversal invariance, the environment time arrow is not transferred to the effective equation of motion. Nevertheless the environment time arrow can be transferred to the solution. The normal mode frequency spectrum is given by the roots of the Fourier transform of the inverse Green function, considered formally without auxiliary conditions for $\tau_s=\tau_e=0$, $D^{-1}_{0,0}(\omega_r)=0$. $\tau_e$ is transferred to the system if the system trajectory, corresponding to a given $\tau_s$ depends on $\tau_e$, namely if
\be
D_{\tau_s,\tau_e}\ne D_{\tau_s,-\tau_e},
\ee
or written equivalently for time reversal invariant full dynamics as
\be
D_{\tau_s,\tau_e}\ne D_{-\tau_s,\tau_e}.
\ee
This may happen that there are normal mode frequencies with imaginary part proportional to $\tau_s\tau_e\epsilon$. These normal modes detect the environment time arrow but this information is not encoded in the effective equation of motion rather than in the auxiliary conditions, represented by the $i\epsilon$ prescription which renders the solution of the equation of motion unique.

In case of a spectrum with condensation point it is more advantageous to use the spectral function
\be
\rho(\Omega)=\sum_n\frac{g_n^2}{2m\omega_n}\delta(\omega_n-\Omega),
\ee
and the self energy
\be\label{selfensi}
\Sigma_{\tau_e}(\omega)=\int d\Omega\frac{2\rho(\Omega)\Omega}{(\omega+i\tau_e\epsilon)^2-\Omega^2}.
\ee
Spectra with condensation point induce nonlocal equation of motion and make the issues of reversibility, causality and runaway modes nontrivial. The integration is passing a pole at infinitesimal distance on the complex $\Omega$-plane and odd powers of $i\omega$ can be generated in the self energy with sign proportional to environment time arrow, $\mr{sign}(\tau_e)$. For instance a simple Ohmic spectral function with smooth suppression at high frequency,
\be\label{ohmspf}
\rho(\Omega)=\frac{\Theta(\Omega)g^2\Omega}{m\Omega_D(\Omega_D^2+\Omega^2)},
\ee
yields 
\be\label{ohmicse}
\Sigma_{\tau_e}(\omega)=-\frac{g^2\pi}{m\Omega_D}\frac{\Omega_D+i\mr{sign}(\tau_e)\omega}{\omega^2+\Omega^2_D},
\ee
and an $\ord{\Omega_D^{-1}}$ Newtonian friction force is found in the effective equation of motion \eq{efseqm} for large cutoff, $\Omega_D\gg\omega_0$. The appearance of $\mr{sign}(\tau_e)$ in $\Im\Sigma_{\tau_e}(\omega)$ indicates irreversibility and the transmutation of the environment time arrow to the system. Furthermore there is no protection against having poles of the retarded Green function on the ``wrong'' sheet even for $\tau_s=\tau_e$ and the effective dynamics might be acausal \cite{caus}. When this happens the real part of the pole represents the frequency scale at which acausality manifests itself. This scale is ${\cal O}(\Omega^{2/3}_D)$ for the Ohmic spectral function and acausality appears at shorter time scale than the characteristic time of the oscillator or the time scale of the friction. It is easy to check that the retarded Green function remains bounded despite the ``wrong'' sign of the imaginary part of the pole.

The retarded Green function of continuous spectrum displays several nontrivial features and calls for a more systematic construction. The quantum version model has been thoroughly analyzed within the CTP formalism \cite{hakim}. We indicate briefly the procedure in the classical description to arrive at a retarded Green function in an unambiguous manner. The classical action is
\be\label{acgen}
S[\hx,\hat y]=\hf\hx\hD_0^{-1}\hx+\hf\sum_n\hat y_n\hat G^{-1}_n\hat y_n+\hx\left(\hj-\sigma \sum_ng\hat y_n\right),
\ee
and the inverse Green functions $\hD_0^{-1}$ and $\hat G^{-1}$ are given by Eq. \eq{hoctpgfinv}, containing the appropriate frequency. The effective action
\be\label{effacgen}
S_{eff}[\hx]=\hf\hx\hD^{-1}\hx+\hj\hx,
\ee
is given in terms of the Green function $\hD=[\hD_0^{-1}-\hat\sigma\hat\Sigma\hat\sigma]^{-1}$ where the self energy, $\hat\Sigma=\sum_ng_n^2\hat G_n$, has the structure of the right hand side of Eq. \eq{trevgrfv}, namely
\bea\label{seflentm}
\Sigma_n(\omega)&=&\frac1m\sum_ng_n^2\frac{\omega^2-\omega^2_n}{(\omega^2-\omega^2_n)^2+\epsilon^2},\nn
\Sigma_f(\omega)&=&-\frac{i\epsilon\tau_e\mr{sign}(\omega)}m\sum_n\frac{g_n^2}{(\omega^2-\omega^2_n)^2+\epsilon^2},\nn
\Sigma_i(\omega)&=&-\frac\epsilon{m}\sum_n\frac{g_n^2}{(\omega^2-\omega^2_n)^2+\epsilon^2}.
\eea

The effective equation of motion will be sought in the parametrization $x_\pm=x\pm x_d/2$ of the CTP trajectories and write the effective action \eq{effacgen} as
\be\label{geffactctp}
S[\hx]=x_d(D^{r-1}x+j),
\ee
where
\be\label{retctpgf}
D_r=\frac1{D_0^{-1}-\Sigma_n-\Sigma_f},
\ee
with $D_0^{-1}=-\partial_t^2-\omega_0^2$. $D_r$ is the retarded system Green function according to Eq. \eq{invprra}. The equation of motion for $x_d$ and $x$ yields $x=-D_rj$ and $x_d=0$, respectively.

\subsection{Dynamical breakdown of time reversal invariance}
Irreversibility, the loss of time reversal invariance of the effective equation of motion by the environment auxiliary conditions is a dynamical symmetry breaking. It is instructive to compare this state of affairs with the spontaneous symmetry breaking of a $Z_2$ symmetry which can be detected in two different manners. One possibility is dynamical, the following of the time dependence of an order parameter and observing its slowing down and becoming frozen at a finite value in the thermodynamical limit. Another, simpler possibility is to inspect the static order parameter, $\Phi$, in the presence of an explicit symmetry breaking term, $j\Phi$, in the Hamiltonian and verify that $\mr{sign}(\Phi)$ is determined by $\mr{sign}(j)$ even when $j$ is infinitesimal. Both signatures of spontaneous breakdown of partial time reversal symmetry can be found in irreversible systems.

The symmetry of the effective equation of motion with respect to time reversal is broken explicitly by the environment initial conditions. These initial conditions are first converted within the environment into an infinitesimally weak dynamical breakdown of symmetry as $t_f-t_i\to\infty$, acting during the time evolution, as pointed out after Eq. \eq{inttctpa}. In the second step, when the effective dynamics is constructed then this explicit, infinitesimal symmetry breaking by $K_f=\ord{\epsilon}$ in the action \eq{ctpfa} gives rise to a finite self energy term $\Sigma_f$ in the effective action \eq{effacgen} when the spectral representation
\be\label{spreprctp}
\hat\Sigma_{\tau_e}(\omega)=\int d\Omega2\Omega\rho(\Omega)\hD(\tau_e\omega,\Omega)
\ee
is used with
\be
\hD(\omega,\Omega)=\frac1m\begin{pmatrix}\frac1{\omega^2-\Omega^2+i\epsilon}&-2\pi i\Theta(-\omega)\delta(\omega^2-\Omega^2)\cr-2\pi i\Theta(\omega)\delta(\omega^2-\Omega^2)&-\frac1{\omega^2-\Omega^2-i\epsilon}\end{pmatrix}.
\ee
cf. \eq{ctpgfho}. The signs of $K_f$ and $\Sigma_f$ are correlated, $\mr{sign}(K_f(\omega))\mr{sign}(\Sigma_f(\omega))=-1$, according to Eq. \eq{invprra}.

The thermodynamical limit, carried out in the dynamical test of spontaneous symmetry breaking is the limit of infinitely many environment degrees of freedom, $N\to\infty$, in our model. The slowing down of an order parameter can be recognized by the non-commutativity of the thermodynamical and the long observational time limits. The toy model with condensation point in its spectrum displays such a non-commutativity. On the one hand, arbitrarily large but finite system appears discrete for $t_f-t_i=\infty$. On the other hand, there is no observation within a finite amount of time which could resolve the spectrum with $N=\infty$ around a condensation point \cite{caus} and we have to rely on the continuous spectrum formalism. The infinitely many normal modes belonging to the unresolved part of the spectrum represent a sink for the system energy and may generate dissipative forces.

\subsection{Causality}
The equations of motion only establish correlation between dynamical quantities at different times without separating cause and effect and the causal relation can be established by the help of the time arrow as mentioned above. The solution of a local equation of motion is always causal because it can be obtained by direct integration where causality is guaranteed. But effective equations of motion are nonlocal and the memory term couples external sources to dynamical quantities of earlier in time, raising the possibility of acausality. It was pointed out after Eq. \eq{grfsw} that the effective dynamics of our toy model with discrete spectrum remains causal but it is easy to see that the self energy \eq{ohmicse} of continuous spectrum has poles with positive imaginary part for certain values of $g$ and $\Omega_D$.

The solution of the equation of motion of a harmonic system, eg. Eq. \eq{efseqm}, can be written as $x=x_{ih}+x_h$, the sum of a particular solution of the inhomogeneous equation, $x_{ih}$, and a general solution of the homogeneous one, $x_h$. The latter belongs to the null-space of the linear equation of motion and is adjusted to satisfy the desired auxiliary conditions. Note that the null-space component of the trajectory drops out from the action thus this adjustment is carried out "by hand", by the choice of an infinitesimal imaginary part of the poles of the Green function, used to obtain $x$. This procedure is replaced in the CTP scheme by a functional definition of the retarded Green function, based on \eq{gffd}, where the imaginary part of the CTP action leads to a well defined pole structure.

The apparent conflict between acausality, arising in case of spectrum with condensation point and the trivially causal trajectory given by the direct integration can be resolved by recalling that any integration quadrature, based on the iteration $t\to t+\Delta t$ of a discretized version of equations \eq{seqm} implies a double limit, namely $N\to\infty$ and $\Delta t\to0$.  The continuum limit, $\Delta t\to0$, carried out first leads to equations which are local in time and therefore causal. If the spectrum has condensation point then we need continuous spectrum and we carry out the thermodynamical limit, $N\to\infty$, first. Let us suppose that we can solve the finite difference equations sequentially, for one coordinate after the other, for a given time. The solution of each second order finite difference equation extends the time interval from which the coordinates are used by $2\Delta t$, backward in time. Therefore such an equation of motion may become nonlocal and develop memory, allowing acausality.

\subsection{Higher order equation of motion and runaway solutions}\label{runtrs}
Let us return first to the effective equation of motion, \eq{efemgm} and assume for the sake of a simple example that the right hand side contains time derivatives of $x$ up to order $n>2$. One might look for the general solution, parametrized by the first $n$ derivatives of the trajectory at the initial time. But this procedure is difficult to implement in realistic situations where one usually knows the initial coordinate and velocity for each degrees of freedom. For instance, the equation of motion, derived for a finite harmonic system with $N<\infty$ in Section \ref{ctpgfvs}, needs $2(N+1)$ auxiliary conditions which can come either from the first $2(N+1)$ derivatives of the system coordinate or from the initial value of the coordinate and velocity for the $N+1$ degrees of freedom. Unless we find a way to transfer the information, residing in the initial coordinates and velocity to the higher order derivative of the system trajectory we can not use the direct integration of the effective equation of motion because that procedure requires the unknown higher order derivatives.

A possible way to find the physically acceptable solutions, satisfying given initial conditions for the environment is to build the information of these environment initial conditions into the scheme in such a manner that two system auxiliary conditions become sufficient to construct a unique system trajectory. This can be achieved by the variational method. The retarded Green function is derived for a harmonic system in Section \ref{ctpgfvs} and it yields a system trajectory which takes into account the environment and satisfies trivial, $x_i=v_i=0$, initial conditions. Non-trivial initial conditions can be introduced by shifting the coordinate and using appropriately chosen external sources. This can easily be shown for a harmonic system, described by the Lagrangian
\be
L(x,\dot x)=\frac{m}2\dot x^2-\frac{m\omega_0^2}2x^2,
\ee
in Eq. \eq{ctplag}, together with the initial conditions $x_i,v_i\ne0$. The action \eq{ctpact} now reads in condensed form as $S_{CTP}[\hat x]=\hf\hx\hD^{-1}\hx+\hx\hj$ and the initial conditions \eq{ctpinc} are used. The shift, $\hx=\hx_i+\hat y$, transforms the action and initial conditions into
\be
S_{CTP}[\hat y]=\hf\hat y\hD^{-1}\hat y+\hat y(\hj+\hD^{-1}\hx_i)+\hj\hx_i
\ee
and $y_i=0$ and $\dot y_i=v_i$. The remaining nontrivial initial condition is taken into account by the modification of the source, $\hj(t)\to\hj(t)+mv_i\delta(t)$, and by placing the trivial initial condition for the velocity slightly before the initial time, $\dot y(t_i-\dt)=0$ where $\dt=0^+$. It is easy to check by integrating the equation of motion for  $t_i-\dt\le t\le t_i+\dt$ that the singularity in the source induces the desired initial velocity but it is more reassuring to see this in a regularized manner, by letting the time advance in finite $\dt$ steps,
\bea
\frac{\hat y(t_i+\dt)-\hat y(t_i)}{\dt}&=&\frac{\hat y(t_i)-\hat y(t_i-\dt)}{\dt}+\frac{\dt}m\left(-m\omega_0^2\hat y(t_i)+\frac1{\dt}m\hat v_i\right)\nn
&=&\hat v_i
\eea
after ignoring terms $\ord{\dt}$ and  $\ord{\epsilon}$. Once the issue of initial conditions is settled then the perturbative, iterative solution can be used for weakly coupled anharmonic systems.

Let us consider this scheme in the context of the toy model, introduced in Section \ref{toymod} where the Ohmic spectral function \eq{ohmspf}, satisfying the stability bound
\eq{stab} produces the simple, local equation of motion,
\be\label{rrem}
0=\gamma\dddot x-\ddot x+\omega_0^2x,
\ee
with general solution,
\be
x(t)=c_+\cos\omega_+t+c_-\cos\omega_-t+c_re^{\omega_rt},
\ee
where $\omega_\pm=\pm\omega_1+i\omega_2$, $\omega_2=\ord{\gamma}$, $\omega_r=+\ord{\gamma^{-1}}$ are the roots of the equation $0=i\gamma\omega^3+\omega^2-\omega_0^2$ and $c_\pm$ and $c_r$ are constant.

It is clear that such a simple model with Hamiltonian bounded from below can not produce runaway trajectories, there ought to be a relation among the physically acceptable values of the parameters which excludes such solutions \cite{dirac}. This restriction is clearly satisfied if we construct the trajectory with the help of the Green function, derived above. The full system and environment starts with trivial initial conditions at $t=-\infty$ in this case and an external source, coupled linearly to the system coordinate drives the system to the desired initial state of motion for $t=t_i>-\infty$. The information, represented by the missing auxiliary conditions of the higher order equation is transferred to the Green function.

One finds in this manner no runaway solutions. In fact, the usual calculation of the retarded Green function, based on the use of the residuum theorem may not always be causal but it remains bounded, $|D(t)|<\infty$. Namely, the solution \eq{phtraj}, written as
\be
x(t)=-\int_{-\infty}^\infty dt'D_r(t,t')j(t')
\ee
in the limit $t_i\to-\infty$, $t_f\to\infty$ contains the retarded Green function given by the residuum theorem in such a manner that the runaway solutions are present for $t'<t$ only. In other words, the normal modes which diverge in the future or in the past are allowed to be present in the solution before or after observation, when they are bounded. What is left from the runaway instability is acausality.

The initial conditions, used in this work, are imposed at early enough, $t_i\to-\infty$, to make the effective equation of motion invariant under translation in time and to be conform with quantum field theory where the application of the Wick theorem requires the adiabatic switching on the interactions. The runaway solutions are removed because the corresponding normal modes are turned off in the retarded Green functions for $t>0$. But the runaway modes may remain suppressed if the initial conditions are set at a final time, $t_i=0$, as well \cite{johnson}. In this case the effective equation of motion is not invariant under time translation but the higher order derivative terms are absent at the initial time and they turn on smoothly. As a result the effective equation of motion needs no additional initial conditions. This case shows nicely that the environment initial conditions make their way into the effective dynamics gradually, in a finite amount of time.

\section{Point charges}\label{pcefths}
Let us now finally consider a system of point charges moving in the presence of an external electromagnetic field $A^e_\mu(x)$, described by the action
\be\label{staaction}
S[x,A]=-\sum_a\int ds\left(m_0c\sqrt{\dot x_a^2(s)}+\frac{e}c\dot x_a^\mu(s) [A^e_\mu(x_a(s))+A_\mu(x_a(s))]\right)+\frac1{16\pi c}\int dxF^{\mu\nu}[1+Q(\Box)]F_{\mu\nu}
\ee
where $x_a^\mu(s)$ denotes the world line of the $a$-th point charge, the parameter $s$ is chosen to be the invariant length after the derivation of the variational equation of motion for the world line, $F_{\mu\nu}=\partial_\mu A_\mu-\partial_\nu A_\mu$ and the dynamics of the EMF is modified by the operator $Q(\Box)$ to be defined below. The effective action for the charges \cite{sctefo},
\be\label{stf}
S_{eff}[x]=-m_0c\sum_a\int ds\left(\sqrt{\dot x_a^2(s)}+\frac{e}c\dot x_a^\mu(s) A^e_\mu(x_a(s))\right)-2\pi\frac{e^2}c\sum_{a,b}\int dsds'\dot x_a(s)D_n(x_a(s)-x_b(s'))\dot x_b(s'),
\ee
involves the symmetric near field Green function and contains no radiation. Note the similarity of the interaction term, the second integral, with Wilson-loops in gauge theories. A possible, non-conventional way to recover it is to assume that all electromagnetic radiations are finally absorbed \cite{wheel}. Instead of following this line of thought we construct the effective theory within the framework of the CTP formalism where the issue of retardation is handled in an automatic manner.

\subsection{CTP effective action}\label{ctpeaed}
The time translation invariant action \eq{inttctpa}, used here for the EMF yields the CTP action
\bea\label{ctpaed}
S_{CTP}[\hx,\hA]&=&-\sum_{\sigma=\pm}\sigma\sum_a\int ds\left(m_0c\sqrt{\dot x_{\sigma,a}^2(s)}+\frac{e}c\dot x_{\sigma,a}^\mu(s)[A^e_{\sigma,\mu}(x^\sigma_a(s))+A_{\sigma,\mu}(x_{\sigma,a}(s))]\right)\nn
&&+\frac1{8\pi c}\int\frac{dp}{(2\pi)^4}(A_{+,\mu}(-p),A_{-,\mu}(-p))\left(g^{\mu\nu}-\frac{p^\mu p^\nu}{p^2}\right)Q(-p^2)\begin{pmatrix}-p^2-i\epsilon&2i\epsilon\Theta(-p^0)\cr2i\epsilon\Theta(k^0)&p^2-i\epsilon\end{pmatrix}\begin{pmatrix}A_{+,\nu}(p)\cr A_{-,\nu}(p)\end{pmatrix}.
\eea
The elimination of the EMF leads to the effective action
\be\label{effactst}
S_{CTPeff}[\hx]=-m_0c\sum_a\int ds\left[\sqrt{\dot x_{+,a}^2(s)}-\sqrt{\dot x_{-,a}^2(s)}\right]+\frac{2\pi e^2}c\sum_{a,b}\int dsds'\dot{\hx}_a(s)\hat\sigma\hD(x_a(s)-x_b(s'))\hat\sigma\dot{\hx}_b(s'),
\ee
where the matrix $\hat\sigma$ is generated by the negative sign on the right hand side of Eq. \eq{ctpact} and the massless CTP Green function is given for $Q=0$ by the Fourier transform of Eq. \eq{emfgfv},
\be\label{rsemgrf}
\Re\hD=\frac{\delta(x^2)}{4\pi}\begin{pmatrix}-1&\mr{sign}(x^0)\cr-\mr{sign}(x^0)&1\end{pmatrix}.
\ee
The effective action is now given by
\bea\label{effc}
S_{CTPeff}[\hx]&=&\hskip-3pt-m_0c\sum_a\int ds\left[\sqrt{\dot x_{+,a}^2(s)}-\sqrt{\dot x_{-,a}^2(s)}\right]\nn
&&+\frac{e^2}{2c}\sum_{a,b}\int dsds'[\dot x_{+,a}\delta((x_{+,a}-x'_{+,b})^2)\dot x'_{+,b}-\dot x_{-,a}\delta((x_{-,a}-x'_{-,b})^2)\dot x'_{-,b}]\nn
&&+\frac{e^2}c\sum_{a,b}\int dsds'\dot x_{+,a}\mr{sign}(x^0_{+,a}-x'^0_{-,b})\delta((x_{+,a}-x'_{-,b})^2)\dot x'_{-,b},
\eea
where the notation $x=x(s)$, $x'=x(s')$ is employed.

The structure of the effective action \eq{effc}, the sum of terms like $e^2\dot x_{\sigma,a}\dot x'_{\sigma',b}/2$ evaluated at light-like separation, shows clearly the way interactions are organized in the CTP scheme. The conserving interactions are represented by single time axis contribution in the second line of eq. \eq{effc}. The summation over $a$ and $b$ symmetrizes these contributions with respect to time inversion and makes up the near field interactions in agreement with the observation that the near field created by a point charge system remains localized and the energy-momentum of the charge plus its near field are preserved.

\begin{figure}
\includegraphics[scale=.6]{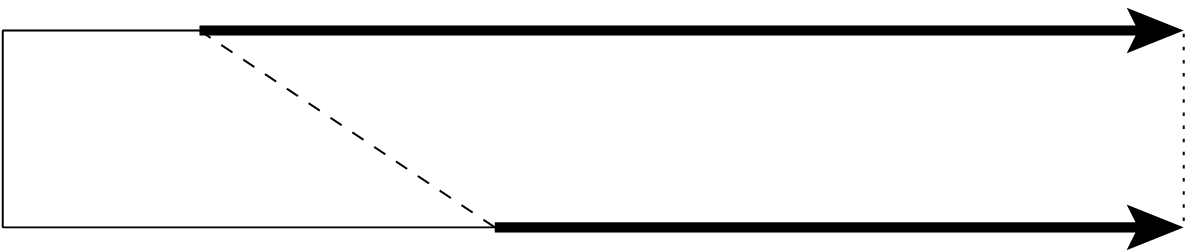}
\centerline{(a)}
\includegraphics[scale=.6]{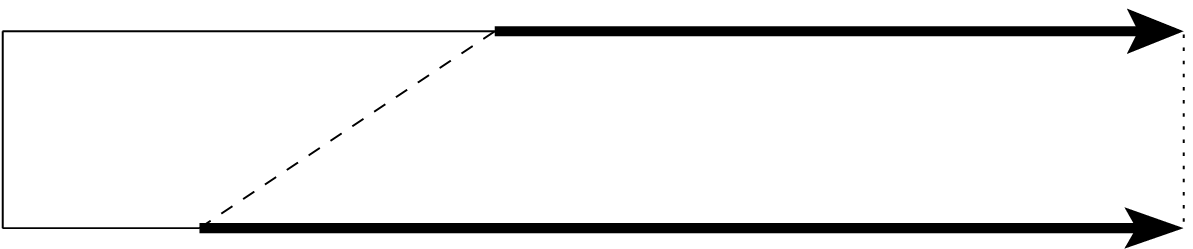}
\centerline{(b)}
\includegraphics[scale=.6]{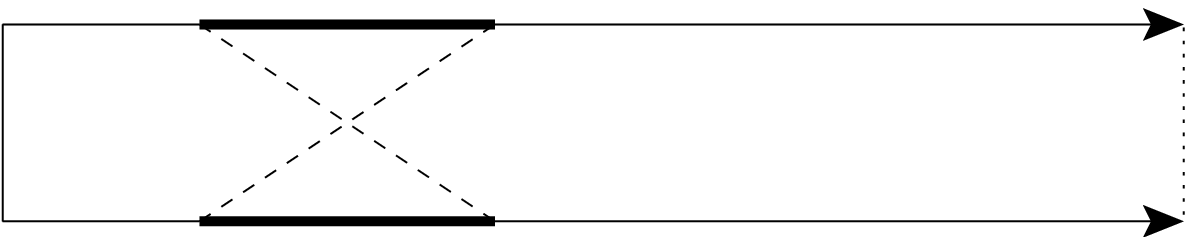}
\centerline{(c)}
\caption{The contribution of the non-conserving interaction to the action, shown in the second line of eq. \eq{effc}. The upper and lower horizontal lines represent the $\sigma=-$ and $\sigma=+$ time axes, respectively. The time $t$ flows from the left to right, from the initial conditions, represented by a vertical line to the final time where the vertical dotted line stands for the final conditions. The dashed lines denote the interactions, mediated by the far field. (a): $x_a^{+0}>x'^{-0}_b$, (b): $x_b^{+0}<x'^{-0}_a$, (c): (a)+(b).}\label{farfield}
\end{figure}

The non-conserving interactions in the third line of \eq{effc} correspond to the far field which decouples from the charge. A contribution with $x_{+,a}^0>x'^0_{-,b}$ is depicted in Fig. \ref{farfield} (a). The retarded EMFs, induced by the charges are present at times $t_+$ and $t_-$ on the $\sigma=+$ and $\sigma=-$ time axes, respectively where $x_a^0<t_+<t_f$ and $x_b^0<t_-<t_f$ and they are matched at $t_+=t_-=t_f$, according to the final condition, eq. \eq{ctpfin}, imposed on the EMF. Their contribution to the influence functional is represented by fat lines in the figure.

The fat lines run until the final time, $t_f$ in Figs. \ref{farfield} (a). This is to draw the attention to a somehow hidden feature of the CTP formalism, namely that contributions to the influence functional, the interactions or the couplings between the time axes extend to a time interval $\tilde t_1<\tilde t<\tilde t_2$ in the notation introduced at the beginning of Section \ref{ctps} and this regime includes the final time, $\tilde t_1<t_f-t_i<\tilde t_2$. Hence the non-conserving interactions originate from processes which are going on at the final time, whatever late is it. Such a long time asymptotic behavior identifies the EMF field component, the radiation field, which generates energy-momentum loss and dissipation for the charges.

Another contribution, obtained by interchanging the charges, $a\leftrightarrow b$, is shown in Fig. \ref{farfield} (b). Due to the relative minus sign between the terms of different time axes in the CTP Lagrangian, cf. eq. \eq{ctpact}, the sum of these two contributions, displayed in Fig. \ref{farfield} (c), seems as if the far field would be emitted by the charge at one moment and would be absorbed by the other charge at another time, just as in QED. The relative minus sign, transferred from the CTP action to the two axis contributions of Fig. \ref{farfield} (c), is the origin of the factor $\mr{sign}(x^0_{+,a}-x'^0_{-,b})$ in the second line of eq. \eq{effc}, representing the antisymmetry of the far field Green function with respect to time reversal.

The self-interaction, the contributions $a=b$ to the effective action \eq{effc} need regularization. An UV regulator has already been proposed to classical electrodynamics \cite{cedreg} but the regulator, applied in this work satisfies more stringent conditions, specified below. The point is the coincidence of the singularities of the distributions $\delta(x^2)$ and $\mr{sign}(x^0)$ which must be shifted away from each other. The problem can be seen by noting that contributions to the effective action, containing $\mr{sign}(x^0)$ are not Lorentz invariant and must be suppressed to recover Lorentz invariant effective dynamics. This is a problem of point charges only because Lorentz invariance remains intact for continuous charge distribution. In fact, the limit $x^0\to0$ appears together with $\v{x}\to0$ in the effective action \eq{effc}, because a pair of events with space-like separation plays no role in classical dynamics ($a\ne b$), and the interaction of continuously distributed charges mediated by the far field is regular, therefore it is negligible in sufficiently small volume ($a=b$).

The regularization $\delta(z)\to\delta_\ell(z)$, $\ell$ being the UV cutoff of length dimension is the same in the near and far Green functions to keep the EMF causal and satisfies the following conditions: the suppression of the divergence, arising from the time derivative acting on $\mr{sign}(x^0)$ requires
\be\label{rege}
\delta_\ell(0)=0.
\ee
Due to Lorentz invariance the Dirac-delta of the far field Green function assumes the same value at space-time events visited by the massless radiation field. This value must obviously be non-vanishing which together with the property \eq{rege} forces us to shift the support of the regulated Dirac-delta slightly off light-cone. To recover the correct energy-momentum flux of the radiation field we need
\be\label{regk}
\int_0^adz\delta_\ell(z)=1
\ee
for any fixed $a>0$ as $\ell\to0$. Lorentz invariance requires the separation of the future and past components of the support of the regulated Green function in such a manner that each of them has a unit weight at any space separation. A simple choice which regulates fully the self interaction is $\delta_\ell(z)=\delta(z-\ell^2)$. But the form
\be\label{reg}
\delta_\ell(z)=\frac{\Theta(z)}{12\ell^4}ze^{-\frac{\sqrt{z}}\ell}
\ee
will be used below because it renders the poles of the retarded Green function a treatable analytical function of the cutoff.

Note that the condition \eq{regk} solves another problem of the self-force: one understands the factor $2\pi$ in front of the second term on the right hand side of Eq. \eq{stf} as $1/2$ times $4\pi$ where the factor half represents the double counting of interactions by the independent summation over $a$ and $b$. But this generates an unwanted factor half for self interaction. This factor is compensated by integrating over both the past and the future light cone in the calculation of the Li\'enard-Wiechert potential of a point charge.

\subsection{Lagrangian of a single charge}
We are interested in the radiation reaction force therefore it is sufficient to consider the effective action \eq{effc} for a single charge. Though the external field $F^e$ is an essential ingredient to induce radiation reaction in case of a single charge it appears in the equation of motion in a trivial manner and will be suppressed. It is advantageous to use the parametrization $x_\pm=x\pm x_d/2$ because we know that $x_d=0$ holds for the solution thus it is sufficient to calculate the effective action up to $\ord{x_d}$. The free action is of the form
\be
S_{free}[x]=-\sum_{\sigma=\pm}\sigma m_0c\int ds\sqrt{\left(\dot x+\frac\sigma2\dot x_d\right)^2}
\ee
and the near and far field contributions to the influence functional are
\bea\label{effacnf}
S_{eff,n}&=&-\frac{e^2}c\int dsds'[\dot x_d\dot x'\delta_\ell(R^2)+\dot x\dot x'\delta'_\ell(R^2)RR_d],\nn
S_{eff,f}&=&-\frac{e^2}c\int dsds'\{2\dot x\dot x'[\mr{sign}(R^0)Rx_d\delta'_\ell(R^2)+x^{d0}\delta(R^0)\delta_\ell(R^2)]-\dot x_d\dot x'\mr{sign}(R^0)\delta_\ell((R^2)\},
\eea
respectively where $R=x'-x$ and $R_d=x'_d-x_d$. The dominant contribution to the integrals comes from $s\sim s'$ and the non-locality of the dynamics is $\ord{\ell}$. We write $\dot x'\approx\dot x+u\ddot x+u^2\dddot x/2$, $R\approx u\dot x+u^2\ddot x/2+u^3\dddot x/6$ and $R^2\approx u^2-u^4\ddot x^2/12$ where $u=s'-s$ and introduce the moments
\be
a_{j,k}=\int_{-\infty}^\infty du\delta_\ell^{(j)}(u^2)|u|^k=c_{j,k}\ell^{k+1-2(j+1)},
\ee
where $\delta_\ell^{(j)}(z)=d^j\delta_\ell(z)/dz^j$. The relation  $a_{j+1,k+2}=-(k+1)a_{j,k}/2$ holds for arbitrary value of $\ell$ and the property $a_{j,2k+1}=\delta_{j,k}(-1)^jj!$ is recovered in the limit $\ell\to0$.

The effective Lagrangian will be calculated in two approximations, first by ignoring $\ord{\ell}$ terms which vanish when the cutoff is removed but keeping any power of $x$, next by skipping $\ord{x^3}$ anharmonic contributions but keeping all cutoff dependences in the quadratic Lagrangian. A straightforward calculation gives the $\ord{x_d}$ Lagrangian
\be\label{feffl}
L_{CTPeff}=x_d\left[m_0c\ddot x+\frac{c_{0,0}e^2}{2c^2\ell}\ddot x-\frac{2e^2}{3c}(\dddot x+\ddot x^2\dot x)\right]
\ee
up to vanishing contribution as $\ell\to0$ after ignoring total $s$-derivatives. The omitted terms which vanish in the limit $\ell\to0$ include higher order derivatives and non-linear combinations of $x$. 

To find the full cutoff-dependence of the quadratic action we return to the influence functional \eq{effacnf} and write the quadratic part of the corresponding Lagrangian as
\be\label{effle}
L_{CTPeff}=x_d\left\{m_0c\ddot x+\frac{4e^2}c\int_{-\infty}^0 du\delta'_\ell(u^2)[x(s+u)-u\dot x(s+u)-x]\right\}
\ee
after ignoring total derivatives and terms $\ord{x^{d2}}$ and $\ord{x^3}$, where $\delta'(a)=d\delta(z)/dz$ and the argument $s$ is suppressed, $x(s)=x$. The cancellation between the near and far components of the  self energy is complete in the future for $\ell\ne0$, leaving behind a memory term and a Volterra-type integro-differential equation of motion. A simple power counting is sufficient to find the behavior in the limit $\ell\to0$, the dominant contribution to the integral comes from $u=\ord{\ell}$ and the order of magnitude $\delta'_\ell(u^2)=\ord{\ell^{-4}}$ shows that $\ord{u^n}$ terms are divergent when $n\le2$, finite for $n=3$ and vanishing for $n\ge4$ as $\ell\to0$. There are no logarithmic corrections for $n=3$ in this classical model. These contributions are separated by writing Eq. \eq{effle} as
\be\label{seplagr}
L_{CTPeff}=x_d\left\{m_0c\ddot x+\frac{c_{0,0}e^2}{2c^2\ell}\ddot x-\frac{2e^2}{3c}\dddot x+\frac{4e^2}cx_d\int_{-\infty}^0 du\delta'_\ell(u^2)\left[x(s+u)-u\dot x(s+u)-x+\frac{u^2}2\ddot x+\frac{u^3}3\dddot x\right]\right\},
\ee
where the square bracket in the integrand approaches zero sufficiently fast at $u=0$ to make the whole integral vanishing as $\ell\to0$.

\subsection{Renormalization}
A divergent, $\ord{\ell^{-1}}$ part of the Lagrangians obtained above can be treated as a mass renormalization in a manner similar to quantum field theories. The renormalization, the removal of the cutoff is carried out by keeping the $\ord{\ddot x}$ term of the Lagrangian cutoff-independent, i.e. by making the bare mass cutoff-dependent,
\be
m_0(\ell)=m-\frac{c_{0,0}e^2}{2c^2\ell},
\ee
where $m$ is the observed, physical mass. The resulting equation of motion,
\be\label{nonveqm}
mc\ddot x=\frac{e}cF^e\dot x+TK
\ee
where the external field $F^e$ is reintroduced again, involves
\be\label{alfeq}
K=\frac{2e^2}{3c}\dddot x-\frac{4e^2}c\int_{-\infty}^0 du\delta'_\ell(u^2)\left[x(s+u)-u\dot x(s+u)-x+\frac{u^2}2\ddot x+\frac{u^3}3\dddot x\right]+\ord{x^2\ell},
\ee
where the projector $T^{\mu\nu}=g^{\mu\nu}-\dot x^\mu\dot x^\nu$, arising from the expansion of the regulated Dirac-delta, projects on the transverse subspace of $\dot x$. The cutoff-independent, $\ord{\ell^0}$ Abraham-Lorentz force comes from the far field component of the self energy, the first term on the right hand side of the second equation of \eq{effacnf} and its sign changes if the time arrow of the EMF is flipped, indicating that this force is generated by the spontaneous breakdown of partial time reversal invariance and represents the time arrow transferred dynamically from the EMF to the charge. The equation of motion \eq{nonveqm} is linear except the projector, which represents a modest partial resummation of the $\ord{x^2\ell}$ terms, arising from the expansion of the regulated Dirac-delta.

Note that the Lagrangian \eq{effle} does not contain the higher order derivative term for $\ell\ne0$. The first term on the right hand side of Eq. \eq{alfeq} survives the limit $\ell\to0$ due to the singularity of the memory kernel, the factor $\delta'(u^2)$ of the integrand. The cutoff-independent Abraham-Lorentz force is generated by the nonuniform convergence of the memory term of the effective Lagrangian in the limit $\ell\to0$. The convergence must be non-uniform because the EMF Green function is a generalized function, a Dirac delta in particular, which can not lead to uniformly convergent integral.

To find physically better motivated parameters we evaluate the Lagrangian \eq{effle} for the world lines $x(s)=x_0e^{-i\omega s}$ and $x_d(s)=x_d$. The choice of the regulated Dirac-delta \eq{reg} which gives $c_{0,0}=1/3$ leads to
\be
L_{CTPeff}=-x_dx_0m_0c\omega^2\left[1+\frac{\lambda_0}6\frac{1+i\ell\omega}{(1-i\ell\omega)^3}\right]
\ee
with $\lambda_0=e^2/m_0c^2\ell$. One can write this expression by absorbing the cutoff-dependence into the renormalized parameters as
\be
L_{CTPeff}=-x_dx_0mc\omega^2\left[1+\lambda\left(\frac23r_0i\omega+r_0\omega\ord{\ell\omega}\right)\right]
\ee
where $m=m_0(1+\lambda_0/6)$ and
\be\label{recc}
\lambda=\frac{\lambda_0}{1+\frac{\lambda_0}6}=\frac{e^2}{mc^2\ell},
\ee
denotes the classical charge radius expressed in units of the UV cutoff. It plays the role of a renormalized strength of self interaction, described by the Abraham-Lorentz force in the renormalized theory. The bare coupling constant,
\be
\lambda_0=\frac\lambda{1-\frac\lambda6},
\ee
as the function of the cutoff possesses a pole in as in the case of the Landau-pole of perturbative QED where the diverging coupling constant is induced at the Landau pole by the one-loop photon self energy. But there is no EMF self-energy for fixed charge world lines and our singularity arises due to the mass renormalization.

\begin{figure}
\includegraphics[scale=.6]{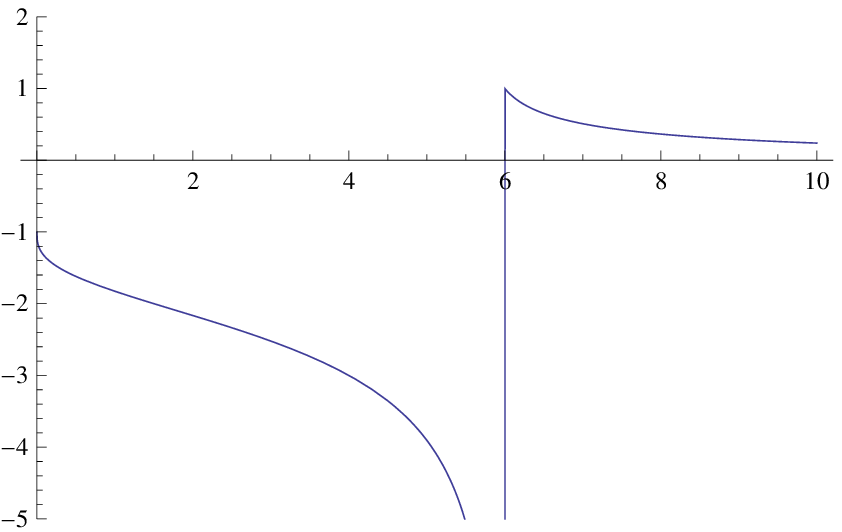}\hskip.5cm\includegraphics[scale=.6]{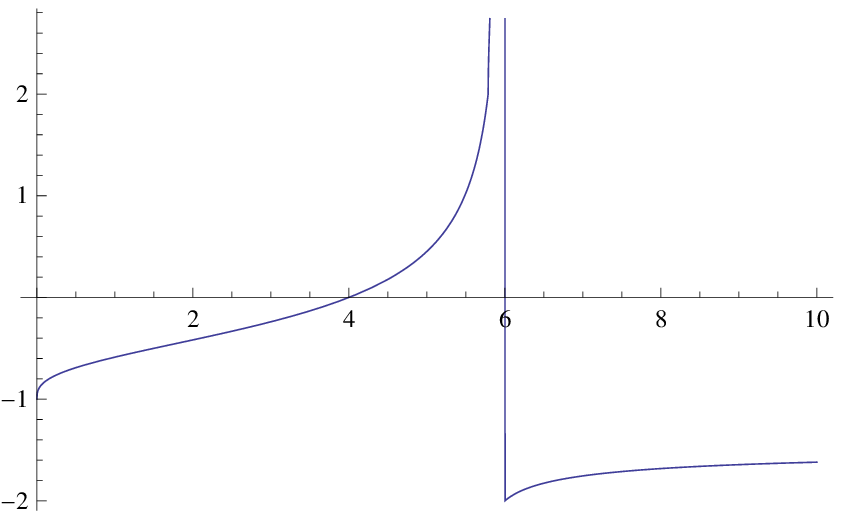}\hskip.5cm\includegraphics[scale=.6]{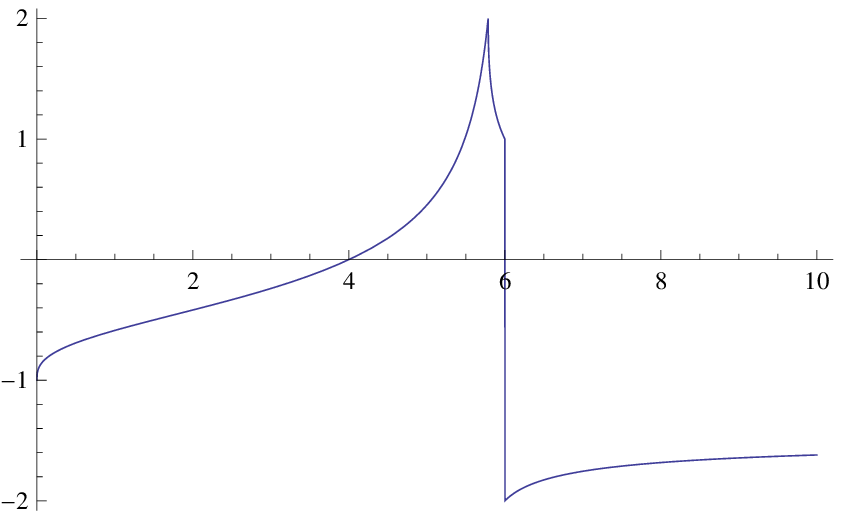}

(a)\hskip5.5cm (b)\hskip5.5cm (c)
\caption{The imaginary part of the poles, (a) $\ell\Im\omega_1$, (b) $\ell\Im\omega_+$ and (c) $\ell\Im\omega_-$ as functions of $\lambda=r_0/\ell$.}\label{impol}
\end{figure}

The parametrization \eq{geffactctp} of the action allows us to read off the retarded Green function of the Lagrangian \eq{effle}, describing the response of a non-relativistic charge with quadratic free Lagrangian on space independent external field,
\be\label{rgfch}
D_r(t)=\int_{-\infty}^\infty\frac{d\omega}{2\pi}\frac{e^{-i\omega t}}{\omega^2+\frac{\lambda_0(\omega^2+i\ell\omega^3)}{6(1-i\omega\ell)^3}}.
\ee
The integrand has five poles, two of them are in the vicinity of zero and give a term proportional to the time as $\ell\to0$, describing the free, causal motion of the charge. The remaining roots are
\bea\label{rgroot}
\omega_1&=&-\frac{i}\ell\left(1+\frac{\lambda_0}{3\sqrt[3]2q}+\frac{q}{3\sqrt[3]4}\right),\nn
\omega_\pm&=&-\frac{i}\ell\left(1-\frac{\lambda_0(1\pm i\sqrt{3})}{6\sqrt[3]2q}-\frac{(1\mp i\sqrt{3})q}{6\sqrt[3]4}\right).
\eea
with $q=\sqrt[3]{\lambda_0(18+\sqrt{324-2\lambda_0})}$. The imaginary parts of the poles are shown in Fig. \ref{impol}. The charge dynamics is causal for $\ell>r_0/4$ but the exact position of the loss of causality naturally depends on the choice of the regulator. A pole is on the ``wrong'' sheet for $\ell<r_0/4$ but the runaway modes do not lead to unstable motion according to Section \ref{runtrs}, the Green function turns the runaway modes off at the time of observation. Both the real and imaginary parts of the poles display a singularity at the classical Landau-pole, $\ell=r_0/6$.

It is natural that a regulated, cutoff theory shows unusual, nonphysical features at the cutoff scale, represented by the modes with finite $\ell\omega$ in our case, e.g. $\omega_\pm=(\pm\sqrt7-3i)/2\ell$ as $\ell\to0$. The reflection of the cutoff-independence of the Abraham-Lorentz force, mentioned above is that the frequency of the mode, representing this force has a cutoff-independent, finite limit,
$\omega_1\to3i/2r_0$ in the renormalized theory. The radiation reaction force extends acausality to length scales shorter than the internal scale, $r_0$.

\subsection{Quantum theory}
We have so far considered an artificial world without quantum effects. These can be handled with the help of the quantum CTP scheme \cite{schw} whose path integral representation consists of integration over the reduplicated CTP doublet trajectories \cite{feynmanv}. The quantum CTP Green function for a harmonic oscillator is identical with the classical Green function, given by Eq. \eq{ctpgfho}. It is illuminating to use the parametrization $x_\pm=x\pm x_d/2$ because quantum fluctuations and decoherence of the coordinate appear in a clearly separable manner through the integration over the trajectories $x(t)$ and $x_d(t)$, respectively.

The generalization of the radiation reaction force for quantized EMF is trivial because the photons are non-interacting in the presence of a fixed charge distribution, their Green function agrees with the classical one, given by Eq. \eq{emfgfv} and the expectation value satisfies the classical Maxwell equation. Hence our previous results apply for this case, as well, for instance Eq. \eq{effle} gives the expectation value of the $\ord{x_d}$ Lagrangian of the influence functional of a single classical charge, averaged over the quantum fluctuations of EMF.

But complications appear in generalizing the radiation reaction force for quantized charge dynamics since there are no trajectories in quantum mechanics. When the definition of the radiation force is sought on the level of observables and their expectation values then the world tube where the charge density expectation value is non-vanishing is always of finite width. As a result, the radiation reaction force depends on the shape of this charge density which is polarized form the vacuum in a complicated manner and the radiation reaction looses its universal form.

Another alternative is to work with the bare effective action, involving an influence functional, depending on the world lines $x_\pm^\mu(\tau)$ of a single charge, still to be integrated over in the path integral. If the decoherence is strong enough to concentrate the CTP path integral over world lines $x^\mu_d=x^\mu_+-x^\mu_-\sim0$ \cite{johnson,galleyh,galleyl} then the Lagrangian \eq{effle} is recovered in $\ord{x_d}$ as the real part of the bare influence functional. In fact, the classical effective action agrees with the quantum one if the latter is calculated with $D_i$ because of the non-interactive nature of photons for fixed charge density, as mentioned above. The imaginary part of the bare influence functional represents a suppressing factor in the integrand $\exp iS$ of the path integral and controls the decoherence of the world line, $x_d^\mu$. The impact of this suppression factor, being quadratic in the electric current, can be recast as a real stochastic Langevin force \cite{feynmanv}, acting on the world line $x^\mu=x_+^\mu-x_-^\mu$. Note that the radiation reaction force of the Lagrangian \eq{effle} applies now for all world lines coming up in the path integral, in particular those with space-like four-velocity \cite{redmount}.

Finally we return to the level of expectation values and note that for a single, strongly decohered and slowly moving point charge the path integration is Gaussian and the Lagrangian \eq{effle} describes correctly the radiation reaction force in terms of the expectation value of the particle trajectory. When the decoherence of the coordinates is weaker or the particle motion is relativistic then the path integration over the world line is not Gaussian anymore and the expectation value of the particle trajectory is subject of a more involved, non-universal radiation reaction force. The structure of quantum corrections in forces due to self interaction or interactions between different particles should be different as decoherence becomes weaker and the motion speeds up.

\section{Classical anomaly}
UV singularities are known to play role in quantum effects of finite scales and this mechanism has a somehow unfortunate name, anomaly, because it was not expected that phenomena with very different scales can be coupled to each other. The cutoff is sent beyond any observation scale in the renormalization procedure and drops out from relations among observed quantities. It was therefore surprising that the presence of an arbitrarily distant cutoff leaves a finite trace in relations among observables. For instance, the rule of partial integration in the action is modified for non-relativistic particles \cite{schulman}. Such an $\ord{\hbar}$ effect of the canonical commutation relation, a quantum anomaly, is due to the nowhere differentiable fractal nature of trajectories which dominate the path integral expressions of non-relativistic quantum mechanics.

The diverging number of one particle states leads further UV singularities in quantum field theories where certain symmetries, such as scale invariance \cite{wilson} or chiral symmetry \cite{abj} are not preserved. Such anomalies result in perturbative quantum field theory from the non-uniform convergence of loop-integrals as the UV cutoff is removed. The order of the integration and the removal of the cutoff can be exchanged without modifying the integral as long as the loop-integral is uniformly convergent. One can first perform the subtraction of the BPHZ renormalization scheme on the integrand itself \cite{collins} in this case, leaving behind no cutoff effects in the renormalized perturbation expansion. But loop-integrals without uniform convergence which are accidentally UV finite and receive no regulating counterterm may converge in a non-uniform manner and retain finite cutoff effects even for arbitrarily large values of the cutoff. Quantum anomalies arise in similar fashion, as well \cite{rgqm}.

Anomalies represent a conflict between the a naive theory without cutoff and the renormalized one where the cutoff is first introduced to render the theory well defined and is next removed. The classical EMF dynamics displays UV and IR anomalies, defined in this sense. The effective dynamics of point charges, outlined above, displays remarkable UV sensitivity which arises from two sources. First, the EMF is divergent at the location of a point charge and second, certain manifestations of the radiation reaction force can be viewed as a magnifying process. In fact, energy-momentum conservation correlates the short distance structure of the EMF around the charge with the radiation field. The latter decouples from its source and carries this information over large distances. Such a sensitivity of the radiation field on the microscopic details of the charge distribution explains the inconsistencies, found in direct calculation of the Lorentz force \cite{la} and its inference by means of energy-momentum conservation \cite{dirac}: the reaction force acquires an artificial multiplicative factor $4/3$ in the limit $\ell\to0$ when the narrow world-tube of the almost point-like charge lacks Lorentz contraction.

The magnification effect of the radiation field stems from the absence of ingoing spherical waves in the radiation field, from the radiation time arrow. This time arrow generates a friction force in the effective charge dynamics, reflecting the dissipation of the energy of the charge to infinitely many IR EMF modes and explains the IR sensitivity of the massless EMF dynamics. In fact, the restriction of the EMF into an arbitrary large but finite volume renders its spectrum discrete and restores reversible effective dynamics.

The UV anomaly is easiest to find in the order of the highest derivative of the world-line in the equation of motion. The integral on the right hand side of Eq. \eq{effle} is the $\ord{e^2\hx^2}$ part of the effective action, it is the retarded part of the one-loop self-energy graph in the perturbative solution of classical electrodynamics and involves the first derivative of the world-line. The non-uniform convergence of this loop-integral which follows from the singular distribution nature of $\delta'(u^2)$ generates a third derivative $\ord{e^2}$ term in Eq. \eq{seplagr} after the removal of the cutoff. As the cutoff-independence of the decay rate $\pi^0\to2\gamma$ is the result chiral anomaly, the cutoff-independent Abraham-Lorentz force reflects the anomalous nature of the classical self-energy graph.

A more direct indication of the non-uniform convergence as the cutoff is removed can be found in the light-cone check, the verification that the radiation field propagates in light-like directions. To test this feature of the EMF we return to the action \eq{staaction} and chose the function $Q$ in such a manner that the propagation of the radiation field is modified as soon as it comes off the light-cone,
\be
\left(D_r\frac1{1+Q}\right)(x,y)=D_{r\ell}(x,y)C\left(\frac{(x-y)^2}{s^2_0}\right)
\ee
where $D_{r\ell}$ is the regulated retarded EMF Green function, $\Box_xD_{r\ell}(x,y)=\delta_\ell(x-y)/4\pi c$ and $C(z)$ is a dimensionless, continuous function satisfying the condition $C(0)=1$. The EMF is off light-cone if the effective charge dynamics depends on the choice of $C(z)$. The finite length parameter, $s_0$ characterizes the scale where the off-shell propagation is tested. Though the renormalized charge dynamics is invariant under such a modification of the EMF dynamics nevertheless the cutoff theory supports no such invariance because the UV regulator places the EMF slightly off-shell due to the relation \eq{regk}. It is obvious that the independence of $Q$ is recovered in the limit $\ell\to0$ for the genuine interaction terms with $a\ne b$ in the effective action \eq{stf} where the distance between the world lines of different point charges is finite and the off-shell effects are of $\ord{\ell/s_0}$. However the $Q$-independence remains a non-trivial issue for the self-interaction, due the UV singular $1/\ell$ factor, emerging from $\delta_\ell(x^2)$ in the EMF Green function.

It is straightforward to rederive the effective Lagrangian for $Q\ne1$. Simple power counting indicates that the effect of an $\ord{z^{n/2}}$ term of $C(z)$ is suppressed in the renormalized theory for $n\ge3$ therefore we choose $C(z)=1+a_1\sqrt{z}+a_2z$, $a_j$ being dimensionless constants. It is easy to see that $a_2$ drops out from the renormalized equation of motion. In fact, $a_2$ can appear in the combination $a_2\dot x/s_0^2$ only on dimensional ground but such a term is forbidden by $\dot x\ddot x=0$. The dependence on $a_1$ survives the removal of the cutoff and the far field contribution to the reaction force modifies the mass,
\be
m\to m+a_1\frac{e^2}{2c^2s_0},
\ee
which can be interpreted as a shift, $1/r_0\to1/r_0+a_1/2s_0$, of the classical charge radius. The UV singular structure, encoded in the space-time derivatives of the far field EMF at the location of the point charge preserves cutoff effects even as $\ell\to0$.

Lorentz symmetry might be broken in classical electrodynamics of point charges, as well. The regulator had to be carefully chosen to avoid the loss of relativistic symmetries. Though not being anomaly, it is reminiscent of the Schwinger term in quantum field theory, the anomalous contribution to the commutator of global Noether currents which arises from the time ordering in the Feynman propagator \cite{treiman}. The anomalous Schwinger term appears in classical field theory, as well, when the commutator is replaced by the Poisson bracket \cite{boulware}. The divergence encountered in this work can be considered as a generalization of such singularities beyond conserved currents.

An IR anomaly, the dependence of the physical content of the theory on the order of the thermodynamical and the long observational time limits, represented by another non-uniform convergence of the loop-integrals affects the status of time reversal invariance. On the one hand, the auxiliary conditions, imposed on the EMF do not violate the invariance of the effective dynamics under time reversal in the presence of an arbitrarily large but finite spatial IR cutoff. On the other hand, dissipative forces arise in a strictly infinite volume. The preservation of a symmetry in finite volume and its violation in the thermodynamical limit is a common feature of models with spontaneous symmetry breaking.

\section{Conclusions}\label{conls}
The effective action was constructed in this work for point charges in classical electrodynamics. To accommodate the dynamical breakdown of time reversal invariance and the resulting dissipative forces in the action principle the CTP formalism was used in classical setting. This scheme (i) distinguishes the non-time reversal invariant contributions coming from the initial conditions and from the equations of motion, (ii) offers a simple view of the origin of irreversibility in effective theories and (iii) separates clearly the near and far field components of EMF. This latter feature leads to two separate classes of the interactions in the effective theory of charges, those which remain localized to the charge system from those which decouple from the charges. Since CTP does not impose constraint on the time evolution and can handle initial condition problems this version of the action principle can easily handle the radiation field, the key ingredient of the reaction force.

The well known Abraham-Lorentz force and the divergent renormalization of the mass have been reproduced and the linearized equation of motion was obtained by retaining the full cutoff dependence. A similar origin of the Abraham-Lorentz force and anomalies has been pointed out, namely these are cutoff-independent effects due to the non-uniform convergence of loop-integrals. The strength of the radiation reaction force displays a pole in its cutoff dependence in a manner reminiscent of the Landau pole.

It is argued that as soon as higher order derivatives appear in an effective equation of motion the solution can not be found by imposing auxiliary conditions, rather one has to use Green functions. When these latter are obtained by the application of the residuum theorem then they always describe stable motion, without runaway solutions. The runaway mode, appearing at $\ell\ll r_0$ leads to acausality at the scale $r_0$.

Phenomena, reminiscent of the UV divergent structure of quantum field theories have been pointed out. The UV singular structure of the retarded far EMF Green function makes the reaction force of point charges dependent on modes of the EMF which propagate off the light-cone even when the UV cutoff is removed. Such a survival of the cutoff effects in the renormalized theory is called anomaly in the quantum case. Another analogy with the UV singular structure of quantum field theory is that a special regularization is required to protect the relativistic symmetries of the self interaction of point charges.

Quantum effects pose further challenges. The classical argument is based on a double limit, both the Planck constant and the size of the charge distribution or the UV cutoff, $\ell$, are sent to zero. These limits do not commute and the wrong order is followed whenever the point charge limit is pursued in classical electrodynamics. The equation of motion, derived above remains the same for the expectation value of the charge coordinate, as far as well decohered, non-relativistic motion is concerned. Therefore true  quantum corrections without classical counterpart show up through world lines with non-monotonous dependence of time along the world line and the interference patterns between neighboring trajectories. It remains to be seen how these effects modify the radiation reaction force.

Extension toward gravity represents a very interesting problem, as well \cite{arnowitt,quinn} where lessons learned from taking into account quantum effects \cite{leonard} can be used to arrive at a unified treatment of the radiation reaction force in Einstein-Maxwell theory.

\acknowledgments
I thank Stanley Deser for bringing Refs. \cite{arnowitt,leonard} to my attention.

\appendix
\section{CTP Green function}\label{ctpgrfcap}
The $2\times2$ CTP block Green function is derived in this Appendix first for a harmonic oscillator \cite{clctp}, a massive scalar field and the EMF.

\subsection{Harmonic oscillator}
The simplest building block in constructing the Green function for free fields is a single harmonic oscillator, described by the action
\be\label{hoaction}
S[x]=\int dt\left(\frac{m}2\dot x^2-\frac{m\Omega^2}2x^2\right).
\ee
We follow its time evolution in discrete steps, by taking $t_j=j\dt-T$, $j=0,\ldots,N$, with $\dt=T/N$. The time interval of the motion is chosen to be $[-T,0]$ in order to remove ambiguities at the final final time in the limit $T\to\infty$. The CTP action \eq{ctpact},
\be
S_{CTP}[x_+,x_-]=\hf\sum_{\sigma,\sigma'}\sum_{\dt-T\le t,t'<0}x_{\sigma,t}\hD^{-1}_{0(\sigma,t),(\sigma',t')}x_{\sigma',t'}+\sum_\sigma\sum_{\dt-T\le t<0}x_{\sigma,t}\hat A_{\sigma,t}z
\ee
where $z=x_0$ and
\bea
\hD^{-1}_{0(\sigma,t),(\sigma',t')}&=&-\delta_{\sigma,\sigma'}\left[\sigma\left(\frac{m}{\dt}(\delta_{t,t'+\dt}+\delta_{t,t'-\dt}-2\delta_{t,t'})+\dt m\Omega^2\delta_{t,t'}\right)-i\dt m\epsilon\delta_{t,t'}\right],\nn
\hat A_{\sigma,t}&=&-\delta_{t,-\dt}\frac{\sigma m}{2\dt}.
\eea
can be written as
\be
S[\hx]=\hf\hx\hD_0^{-1}\hx+\hx\hat Az.
\ee

The calculation of the Green function, the quadratic form of the effective action for $\hat x$ can be carried out by finding the functional $W[\hj]=S[\hx]+\hx\hj,$ cf. Eq. \eq{genfcgfv}. First we eliminate $\hx$ by means of its equation of motion, $\hx=-\hD_0(\hat Az+\hj)$, yielding
\be
W[\hj]=-\hf z\hat A\hD_0\hat Az-z\hat A\hD_0\hj-\hf\hj\hD_0\hj.
\ee
Next we eliminate $z$ by means of its equation of motion, $z=-\hat A\hD_0\hj/\hat A\hD_0\hat A$, to find
\be
W[\hj]=-\hf\hj\hD\hj
\ee
with
\be\label{ctpgfvsta}
\hD=\hD_0-\hD_0\hat A\frac1{\hat A\hD_0\hat A}\hat A\hD_0.
\ee

The detailed expression for this Green function comes from the representation
\be
x(t)=\sqrt{\frac2{T}}\sum_{n=0}^N\tilde x_n\sin\omega_nt
\ee
of discrete trajectories $x(-T)=x(0)=0$ defined at $t=j\dt$, $j=-N,\ldots,-1$ where $\omega_n=\frac\pi{T}n$. For the construction of the Green function \eq{ctpgfvsta} we need
\be
D_0(t,t')=\frac2{Tm}\sum_{n=1}^N\frac{\sin\frac\pi{T}nt\sin\frac\pi{T}nt'}{\frac4{\dt^2}\sin^2\pi\frac{\dt n}{2T}-\Omega^2+i\epsilon},
\ee
and its special values,
\be
D_0(t,-\dt)=-\frac{2}{Tm}\sum_{n=1}^N\frac{\sin\frac\pi{T}nt\sin\frac\pi{T}n\dt}{(\frac\pi{T}n)^2-\Omega^2+i\epsilon},
\ee
and
\be
D_0(-\dt,-\dt)=\frac{2}{Tm}\sum_{n=1}^N\frac{\sin^2\frac\pi{T}n\dt}{\frac{4}{\dt^2}\sin^2\pi\frac{\dt n}{2T}-\Omega^2+i\epsilon}.
\ee
We remove first the UV cutoff by carrying out the limit $N\to\infty$ which is not uniform owing to the non-differentiability of the CTP trajectory at the final time. This is followed by the removal of the IR cutoff, $T\to\infty$, yielding
\bea
D_0(t,t')&=&\frac1{2\pi m}\int_{-\infty}^\infty d\omega\frac{e^{i\omega(t-t')}-e^{i\omega(t+t')}}{\omega^2-\Omega^2+i\epsilon}\nn
&=&-\frac{i}{2m\Omega}\left[e^{-(i\Omega+\frac{\epsilon}{2\Omega})|t-t'|}-e^{(i\Omega+\frac{\epsilon}{2\Omega})(t+t')}\right],
\eea
for $t,t'<0$, together with
\bea
D_0(t,-\dt)&=&\frac{i\dt}{\pi m}\int_{-\infty}^\infty d\omega\frac{\omega e^{i\omega t}}{\omega^2-\Omega^2+i\epsilon}\nn
&=&-\frac{\dt}{m}e^{(i\Omega+\frac{\epsilon}{2\Omega})t}
\eea
and
\bea
D_0(-\dt,-\dt)&=&\frac{2\dt}{m\pi}\int_0^\pi d\omega\left(1-\sin^2\frac{\omega}{2}\right)+\frac{2\dt^2\Omega^2}{m\pi}\int_0^\infty d\omega\frac1{\omega^2-\Omega^2+i\epsilon}\nn
&=&\frac\dt{m}-i\frac{\dt^2\Omega}{m}.
\eea
The initial condition decouples after sufficiently long time evolution if some relaxation is present but the final condition on the CTP trajectories remains always relevant, motivating the use of the time interval $-T<t<0$. The full CTP propagator, given by \eq{ctpgfvsta},
\be
\hD(t,t')=-\frac{i}{2m\Omega}\begin{pmatrix}e^{-i\Omega|t-t'|}&e^{i\Omega(t-t')}\cr e^{-i\Omega(t-t')}&e^{i\Omega|t-t'|}\end{pmatrix},
\ee
recovers translation invariance in time. Its Fourier transform,
\be
\hD(\omega)=\int dte^{i\omega t}\hD(t,0),
\ee
turns out to be
\be\label{ctpgfho}
\hD(\omega)=\frac1m\begin{pmatrix}\frac1{\omega^2-\Omega^2+i\epsilon}&-2\pi i\Theta(-\omega)\delta(\omega^2-\Omega^2)\cr
-2\pi i\Theta(\omega)\delta(\omega^2-\Omega^2)&-\frac1{\omega^2-\Omega^2-i\epsilon}\end{pmatrix}.
\ee
 
The inverse Green function can be obtained by means of the representation $\delta_\epsilon(\omega)=\epsilon/\pi(\omega^2+\epsilon^2)$ of the Dirac-delta in Eqs. \eq{invprra}-\eq{invpri},
\be\label{hoctpgfinv}
\hD_0^{-1}(\omega)=m\hat\sigma\left[(\omega^2-\Omega^2)\begin{pmatrix}1&0\cr0&-1\end{pmatrix}
+i\epsilon\begin{pmatrix}1&2\Theta(-\omega)\cr2\Theta(\omega)&1\end{pmatrix}\right]\hat\sigma,
\ee
or
\be\label{hoctpgfinvc}
D_0^{-1n}=\omega^2-\Omega^2,~~~D_0^{-1f}=i\mr{sign}(\omega)\epsilon,~~~D_0^{-1i}=\epsilon.
\ee

\subsection{Scalar field}
The Green function can easily be found for free field in a similar manner. The expression
\eq{ctpgfvsta} for the Green function remains valid with
\be\label{hoctpactk}
\hat A_{\sigma,t}(\v{p})=-\delta_{t,-\dt}\frac\sigma{2\dt}
\ee
given for the Fourier mode $\v{p}$ and 
\be\label{dnullsa}
D_0((t,\v{x}),(t',\v{x}'))=\frac2{Tm}\int\frac{d^3p}{(2\pi)^3}e^{i\v{p}(\v{x}-\v{x}')}\sum_{n=1}^N\frac{\sin\frac\pi{T}nt\sin\frac\pi{T}nt'}{\frac4{\dt^2}\sin^2\pi\frac{\dt n}{2T}-\Omega^2(\v{p})+i\epsilon},
\ee
where $\Omega(\v{p})=\sqrt{m^2+\v{p}^2}$ for a scalar field of mass $m$, in units $c=\hbar=1$. The limit $T\to\infty$ gives for the Fourier transform
\be
\hD(p)=\int d^4xe^{ipx}\hD(x)
\ee
the result
\be\label{freectpgfv}
\hD(p)=\begin{pmatrix}\frac1{p^2-m^2+i\epsilon}&-2\pi i\Theta(-p^0)\delta(p^2-m^2)\cr
-2\pi i\Theta(p^0)\delta(p^2-m^2)&-\frac1{p^2-m^2-i\epsilon}\end{pmatrix}.
\ee
The inverse Green function can be written as
\be\label{ifreectpgfv}
\hD^{-1}=(p^2-m^2)\begin{pmatrix}1&0\cr0&-1\end{pmatrix}
+i\epsilon\begin{pmatrix}1&-2\Theta(-p^0)\cr-2\Theta(p^0)&1\end{pmatrix}.
\ee

\subsection{Electromagnetic field}
The transverse part of the EMF Green function is given by a scalar massless Green function up to a sign,
\be\label{emfgfv}
\hD^{\mu\nu}(p)=-\left(g^{\mu\nu}-\frac{p^\mu p^\nu}{p^2}\right)\begin{pmatrix}\frac1{p^2+i\epsilon}&-2\pi i\Theta(-p^0)\delta(p^2)\cr
-2\pi i\Theta(p^0)\delta(p^2)&-\frac1{p^2-i\epsilon}\end{pmatrix}.
\ee
The longitudinal part depends on the gauge fixing and drops out from the effective action due to current conservation.

\end{document}